\documentclass[12pt]{article}
\usepackage{scicite}
\usepackage[normalem]{ulem}
\usepackage{times, color, verbatim }

\usepackage[labelfont=bf]{caption}

\newenvironment{sciabstract}{%
\begin{quote} \bf}
{\end{quote}}

\newcounter{lastnote}

\usepackage{graphicx,amsmath,float,array}
\usepackage[in,empty]{fullpage}

\newcommand{\br}{\ensuremath{\hat{b}^{\dagger}}}
\newcommand{\bl}{\ensuremath{\hat{b}}}

\newcommand{\dr}{\ensuremath{\hat{d}^{\dagger}}}

\newcommand{\xzp}{\ensuremath{x_{zp}}}
\newcommand{\xzpsq}{\ensuremath{x_{zp}^2}}

\newcommand{\XI}{\ensuremath{\hat{X}_1}}
\newcommand{\XIsq}{\ensuremath{\langle\Delta\hat{X}_1^2\rangle}}
\newcommand{\XII}{\ensuremath{\hat{X}_2}}
\newcommand{\XIIsq}{\ensuremath{\langle\Delta\hat{X}_2^2\rangle}}

\newcommand{\w}{\ensuremath{\omega}}

\newcommand{\wm}{\ensuremath{\omega_m}}
\newcommand{\wc}{\ensuremath{\omega_c}}
\newcommand{\gm}{\ensuremath{\gamma_m}}

\newcommand{\kc}{\ensuremath{\kappa}}

\newcommand{\kR}{\ensuremath{\kappa_{out}}}

\newcommand{\go}{\ensuremath{g_0}}
\newcommand{\wred}{\ensuremath{\omega_-}}
\newcommand{\wblue}{\ensuremath{\omega_+}}

\newcommand{\Geff}{\ensuremath{\mathcal{G}}}
\newcommand{\Gr}{\ensuremath{G_-}}
\newcommand{\Gb}{\ensuremath{G_+}}

\newcommand{\npr}{\ensuremath{n_p^-}}
\newcommand{\npb}{\ensuremath{n_p^+}}

\newcommand{\ncth}{\ensuremath{n_c^{th}}}

\newcommand{\nmth}{\ensuremath{n_m^{th}}}

\author
{E. E. Wollman,$^{1\ast}$ C. U. Lei,$^{1\ast}$ A. J. Weinstein,$^{1\ast}$ J. Suh,$^{2}$ A. Kronwald,$^{3}$\\ F. Marquardt,$^{3,4}$ A. A. Clerk,$^{5}$ K. C. Schwab$^{1\dagger}$\\
\\
\normalsize{$^{1}$Applied Physics, California Institute of Technology, Pasadena, CA 91125, USA}\\
\normalsize{$^{2}$Korea Research Institute of Standards and Science, Daejeon 305-340, Republic of Korea}\\
\normalsize{$^{3}$Institute for Theoretical Physics, Friedrich-Alexander-Universit{\"a}t} \\\normalsize{Erlangen-N{\"u}rnberg, Staudtstr. 7, 91058 Erlangen, Germany}\\
\normalsize{$^{4}$Max Planck Institute for the Science of Light,}\\ \normalsize{G{\"u}nther-Scharowsky-Stra{\ss}e 1, 91058 Erlangen, Germany}\\
\normalsize{$^{5}$Department of Physics, McGill University, Montreal, Quebec, H3A 2T8 Canada}\\
\normalsize{$^\ast$These authors contributed equally to this work.}\\
\normalsize{$^\dagger$Corresponding author. Email: schwab@caltech.edu.}
}

\date{}

\title{Quantum squeezing of motion in a mechanical resonator}

\begin{document}

\baselineskip24pt


\maketitle

\begin{sciabstract}
As a result of the quantum, wave-like nature of the physical world, a harmonic oscillator can never be completely at rest.  Even in the quantum ground state, its position will always have fluctuations with variance $\mathbf{\Delta \hat{x}_{zp}^2 = \hbar/\left( 2 m \wm\right)}$, called the zero-point motion. This physical theory, which originated over 100 years ago to describe the observed behavior of fundamental particles like photons, electrons, and atoms, has also recently been shown to describe the motion of micron-scale mechanical systems composed of many atoms\cite{OConnell:2010,Weinstein:2014}. Although the zero-point fluctuations are unavoidable, they can be manipulated. In this work, using microwave frequency radiation pressure, we both prepare a micron-scale mechanical system in a state near the quantum ground state and then manipulate its thermal fluctuations to produce a stationary, quadrature-squeezed state. We deduce that the variance of one motional quadrature is 0.80 times the zero-point level, or 1 dB of sub-zero-point squeezing. This work is relevant to the quantum engineering of states of matter at large length scales, the study of decoherence of large quantum systems\cite{Hu:1993}, and for the realization of ultra-sensitive sensing of force and motion.
\end{sciabstract}

\maketitle

In the quantum ground state, a mechanical resonator has position fluctuations divided equally between its two motional quadratures, $\XI$ and $\XII$, which are related to the position operator by: $\hat{x} = \XI \cos{\wm t}+\XII \sin{\wm t}$. The ground-state fluctuations minimize the uncertainty relation given by the quadratures' non-zero commutator: $\langle \Delta \XI^2\rangle \langle \Delta \XII^2\rangle \geq \frac{1}{4} \left| \langle [\XI,\XII] \rangle \right|^2 = \left(\hbar/2 m \wm \right)^2$. Given this uncertainty relation, it is, in principle, possible to squeeze the zero-point noise such that fluctuations in one quadrature are reduced below the zero-point level at the expense of increasing noise in the orthogonal quadrature. More generally, other non-commuting observable pairs can be squeezed, and squeezed states have been created and detected in such varied systems as optical\cite{Slusher:1985} and microwave\cite{Yurke:1988} modes, the motion of trapped ions\cite{Meekhof:1996}, and spin states in an ensemble of cold atoms\cite{Hald:1999}. Transient quantum squeezing has also been created and observed in the motion of molecular nuclei\cite{Dunn:1993} and of terahertz-frequency phonons in an atomic lattice on picosecond timescales\cite{Garrett:1997}. In this work, we demonstrate steady-state squeezing of a micron-scale mechanical resonator by implementing a reservoir-engineering scheme proposed by Kronwald, Marquardt, and Clerk\cite{Kronwald:2013}. This method is closely related to the approach of Cirac et. al.\cite{Cirac:1993} that was recently used to produce quantum squeezed states in the motion of trapped ions.\cite{Kienzler:2015}

A major challenge for quantum squeezing of a radio-frequency mechanical mode is that, even at a temperature of 10 mK, the thermal occupation and corresponding position fluctuations are far larger than the quantum zero-point fluctuations: $\Delta \hat{x}^2 \sim 100\cdot\Delta \hat{x}_{zp}^2$ for a 4 MHz resonator; quantum squeezing\footnote{In this work, we use ``quantum squeezing'' to refer to any quadrature-squeezed state with fluctuations below the zero-point level in one quadrature.} can only be accomplished by first overcoming this large thermal contribution. In contrast, optical modes are found in the quantum ground state at room temperature. Squeezing of mechanical fluctuations was first demonstrated far outside the quantum regime by parametrically modulating the mechanical spring constant\cite{Rugar:1991}. Since parametric methods are limited to 3 dB of steady-state squeezing, the occupation factor of the mechanical mode must be well below one phonon to achieve squeezing below the zero-point fluctuations. There are many theoretical proposals for surpassing the 3 dB limit to produce quantum squeezing\cite{Rabl:2004,Ruskov:2005,Clerk:2008,Zhang:2009,Jahne:2009,Mari:2009,Szorkovsky:2011,Vanner:2011,Gu:2013,Asjad:2014,Lu:2015}, and improvement over the 3 dB limit has been realized experimentally with modified parametric techniques\cite{Szorkovsky:2013,Vinante:2013,Pontin:2014}. Squeezing below the zero-point fluctuations, however, has yet to be achieved.

Squeezing via reservoir engineering, as described in \cite{Cirac:1993}~and \cite{Kronwald:2013}, has advantages over other methods, as it creates a system in which the mechanics relaxes into a steady-state squeezed state without the fast measurements and control necessary for feedback. The scheme consists of applying two pump tones to an optical or microwave cavity parametrically-coupled to a mechanical resonator. The pumps are detuned from the cavity frequency, \wc, by the mechanical frequency, $\pm\wm$, and the red-detuned pump has an intracavity occupation, $\npr$, greater than that of the blue-detuned pump, $\npb$ (Fig. 1A). This is a similar set-up to one used for a back-action-evading (BAE) measurement of a single quadrature\cite{Braginsky:1980}, but with excess red power. In the absence of any blue-detuned power, the pump configuration becomes a single, red-detuned drive, as in sideband cooling\cite{Marquardt:2007}. In this limit, the fluctuations of both quadratures are damped and cooled, but the final noise in each quadrature is limited by the zero-point noise of the cavity and drive. In the other limit of equal red- and blue-detuned power, the pump configuration becomes the balanced drives of BAE. In this limit, all back-action noise is added to the \XII~quadrature, leaving the \XI~quadrature unperturbed by the measurement, and neither quadrature is cooled. In between these limits, both quadratures are damped by the excess red power, while the backaction noise added to \XI~is less than the zero-point noise associated with the total damping. By optimizing the pump ratio between these two limits, it is possible in principle to produce arbitrarily-large amounts of sub-zero-point squeezing (i.e., $>$ 3 dB) if the coupling between the mechanics and the squeezed reservoir sufficiently dominates the mechanical dissipation rate \cite{Kronwald:2013}.

This interplay between damping and back-action can be seen in the equations for the quadrature fluctuations of the system described above. Here, we take the cavity (mechanics) to be coupled to a bath with thermal occupation \ncth~(\nmth) at a rate \kc~(\gm). For optical systems, \ncth~is usually indistinguishable from 0, but for microwave systems, a non-zero \ncth~is commonly observed at high pump powers\cite{Gao:2008,Suh:2013}. The cavity and mechanics are coupled with a single-photon coupling rate $\go=\frac{d\wc}{dx} \xzp$. We also define the enhanced optomechanical coupling rates for the red and blue pumps as $G_\pm = \go \sqrt{n_{p}^\pm}$ and the effective optomechanical coupling rate as $\Geff=\sqrt{\Gr^2-\Gb^2}$. The linearized interaction Hamiltonian for this system is given by
\begin{equation}\label{eq:hamiltonian}
		\hat{H}= -\hbar \dr(\Gb \br + \Gr \bl) -\hbar \dr(\Gb \bl e^{-2 i \wm t} + \Gr \br e^{2 i \wm t}) + \mathrm{H.c},
	\end{equation}
where \dr~is the cavity photon creation operator and \br~is the mechanical phonon creation operator. In the good-cavity limit ($\wm \gg \kc$), and when $\kc \gg \gm$, the quadrature fluctuations are then given by
\begin{equation}\label{eq:x1sq_0detune}
		\XIsq = \xzpsq \bigg\{ \frac{\gm}{\kc} \frac{\left(4\Geff^2+\kc^2\right)}{\left(4 \Geff^2 +\gm \kc\right)} \left( 2 \nmth + 1\right) + \frac{4(\Gr-\Gb)^2}{\left(4 \Geff^2 +\gm \kc\right)} \left( 2 \ncth + 1 \right) \bigg\}.
	\end{equation}
    \begin{equation}\label{eq:x2sq_0detune}
		\XIIsq = \xzpsq \bigg\{ \frac{\gm}{\kc} \frac{\left(4\Geff^2+\kc^2\right)}{\left(4 \Geff^2 +\gm \kc\right)} \left( 2 \nmth + 1\right) + \frac{4(\Gr+\Gb)^2}{\left(4 \Geff^2 +\gm \kc\right)} \left( 2 \ncth + 1 \right) \bigg\}.
	\end{equation}
In both cases, the first term is proportional to $2\nmth+1$ and has a prefactor that is less than 1 for all $\Geff>0$. This term represents the damping of both quadratures due to the excess red power. The second term is proportional to $2\ncth+1$, and is due to the back-action from the microwave field. We see that the back-action is reduced for \XIsq~and increased for \XIIsq~relative to $\frac{4\Geff^2}{\left(4 \Geff^2 +\gm \kc\right)} \left( 2 \ncth + 1 \right)$, the amount of back-action we would normally associate with the net damping rate set by $\Geff^2$. This reduction for \XI~makes it possible to reduce \XIsq~below \xzpsq. 

It is important to note that the quadrature variances depend on the thermal occupations of both the mechanical and cavity baths, which can both be heated by the applied pump power. These thermal occupations are also evident in the noise spectrum of photons leaving the cavity. For our two-port device measured in transmission, the output spectrum derived from the Hamiltonian in Eq. \ref{eq:hamiltonian} has the form
\begin{equation}\label{eq:outspec}
	\bar{S}_{VV}[\w] = S_0 + \kR \kc \frac{(\gm/2)^2+(\w-\wc)^2}{|f[\w-\wc]|^2} \ncth + \kR \gm \frac{\Gr^2 \nmth +\Gb^2(\nmth+1)}{|f[\w-\wc]|^2},
	\end{equation}
where $f[\w]=\Geff^2+(\gm/2-i\w)(\kc/2-i\w)$, \kR~is the coupling through the output port of our device, and $S_0$ is the noise floor of our measurement system. To determine the state of the mechanics, we follow an approach analogous to that used for mechanical cooling measurements in the presence of heated baths\cite{Rocheleau:2010,Teufel:2011,Massel:2012}: we fit the output noise spectrum in the presence of the squeezing tones to extract the bath occupations, and then infer the quadrature variances for a mechanical resonator coupled to these baths.

Our device consists of a mechanical resonator of frequency $\wm=2\pi \times3.6$ MHz coupled capacitively to a microwave lumped-element resonator with frequency $\wc=2\pi \times 6.23$ GHz (Fig. 1B). The bare mechanical linewidth is $\gm = 2\pi\times 3$ Hz at 10 mK, and the cavity linewidth is $\kc = 2\pi \times 450$ kHz. The optomechanical coupling rate is $\go = 2\pi \times 36$ Hz, and the zero-point fluctuations have an amplitude of $\xzp \sim 2.3$ fm. Our measurement set-up is shown in Fig. 1C. We can measure either the output noise spectrum of the system (spectral response), or we can sweep a small probe tone through the cavity to measure the complex transmission (driven response). We first calibrate the normalized spectral power at the output of our system using known thermal occupations to find \go~(Fig. 2A). We then cool our sample temperature to 10 mK, giving an initial thermal occupation of $\nmth \sim 50$ quanta. We calibrate the enhanced optomechanical coupling rate, \Gr, to the output power of a red-detuned pump by measuring the total mechanical linewidth, $\gamma_{tot} = \gm+4 \Gr^2/\kc$, from the driven response of the mechanics for linewidths much smaller than \kc~(Fig. 2B). As we increase the pump power further, the mechanical linewidth becomes comparable to the cavity bandwidth, and a strong-coupling model of the system is needed to fit the driven response. Fits with this model are in excellent agreement to the low-power calibrations, so we see that the device is well behaved and performs linearly over a broad range of pump powers (Fig. 2B). This agreement also confirms that the dynamics of our system are well-described by the linearized Hamiltonian in Eq. \ref{eq:hamiltonian}~in the limit of no blue-detuned power. As we increase the red-detuned pump power, the mechanics become more strongly coupled to the cold cavity bath, reaching a minimum occupation of $0.22 \pm 0.08$ quanta (Fig. 2C). The blue-detuned coupling rate, $\Gb$, is also calibrated to the measured output power using the $\Gr$ calibration and a measurement of the output gain at $\wblue$ vs. $\wred$. For further discussion of the system calibrations, see the supplementary online text.

To squeeze the mechanical motion, a red-detuned pump is applied at $\omega_c-\omega_m$ and a weaker, blue-detuned pump is applied at $\omega_c+\omega_m$ such that their sidebands overlap at the cavity frequency, as described above. We then measure the output spectrum in the lab frame at different pump power ratios and fit a spectral model to the background-subtracted output spectrum normalized by the measured pump power (Fig. 3A). The spectral model includes ``bad-cavity" effects that arise when $\wm \not\gg \kc$, and does not assume that the pumps are aligned at precisely $\wc \pm \wm$. In the absence of these effects, the model reduces to Eq. \ref{eq:outspec}. Since we determine \Gr~and \Gb~from our calibrations, and  measure \kc~to high precision with a driven response, the only unknown parameters are \ncth~and \gm\nmth. For parameter estimation, we utilize a Bayesian analysis that systematically incorporates the uncertainty from both the nonlinear fitting and the uncertainty in the independent calibration measurements. Given the observed data and our knowledge of the system Hamiltonian, we directly sample the posterior probability distribution of all system parameters. We then use this distribution to extract estimators for the bath occupations, \ncth~and \gm\nmth, and for the mechanical quadrature occupations, \XIsq~and \XIIsq~(see the supplementary online text for further information). The extracted bath occupations are plotted in Fig. 3C, and the quadrature variances are shown in Fig. 3D. We find that, at our lowest point, we are squeezed to $0.80 \pm 0.03\, \xzpsq$, or $1.0 \pm 0.2$ dB below the zero-point fluctuations when $\npr = 1.26\times10^7$ and $\npb = 0.51\times10^7$. The spectral fit for this point is shown in Fig. 3B.

To directly measure the fluctuations of a single quadrature, it is possible to introduce a set of weak back-action evading tones in addition to the squeezing tones\cite{Suh:2014} (Fig. 4A). In order for the presence of the probe tones not to interfere with the measured mechanical quadrature, the probe sideband must be separated from the squeezing sideband by many mechanical linewidths. For this device, where the mechanical linewidth at optimal squeezing is comparable to the cavity linewidth, such a separation would place the probe sideband outside the cavity bandwidth, significantly decreasing the probe gain and leading to unrealistic measurement times. However, we have implemented such a measurement scheme using a similar device with a larger cavity bandwidth, $\kc = 2\pi \times 860$ kHz, and narrower mechanical linewidth $\gamma_{tot} = 10$ kHz (Fig. 4B). As is shown in Fig. 4C, the phase-dependent BAE measurement shows the expected squeezed thermal state produced by the two squeezing pumps, with minimum fluctuations of $1.09 \pm 0.05 \, \xzpsq$, limited in a similar way as the first device by heating.  This measurement demonstrates the quadrature squeezing produced by imbalanced pumps at $\wc \pm \wm$ as described by the theory.

While a squeezed thermal state always has a positive Wigner function, when the fluctuations in one quadrature are reduced below the zero-point level, the squeezed state no longer has a well-behaved P-representation -- that is, it cannot be represented as an incoherent mixture of coherent states\cite{Kim:1989}. For this reason, a quantum squeezed state is considered a non-classical state\cite{Gerry:2004}. It is a current goal to study the non-classical behavior in larger and larger systems, and recent progress in the field of opto- and electromechanics has resulted in the generation of mechanical Fock states\cite{OConnell:2010} and entanglement \cite{Palomaki:2013}; quantum squeezing in a micron-scale mechanical resonator is  an important addition to this short list. Future work could study the quantum decoherence of a mechanical object through the decay of the squeezed state\cite{Hu:1993}.

\bibliographystyle{Science}
\bibliography{squeeze}

\subsection*{Acknowledgments}
This work is supported by funding provided by the
Institute for Quantum Information and Matter, an NSF Physics
Frontiers Center with support of the Gordon and Betty Moore
Foundation (NSF-IQIM 1125565), by the Defense Advanced
Research Projects Agency (DARPA-QUANTUM HR0011-10-1-0066), by the NSF (NSF-DMR 1052647 and NSF-EEC 0832819), and by the Semiconductor Research Corporation (SRC) and Defense Advanced Research Project Agency (DARPA) through STARnet Center for Function Accelerated nanoMaterial Engineering (FAME).
A.C., F.M., and A.K. acknowledge support from the DARPA ORCHID program through a grant from AFOSR, F.M. and A.K. from ITN cQOM and the ERC OPTOMECH, and A.C. from NSERC.

\begin{figure*}[p]
\begin{center}
\includegraphics{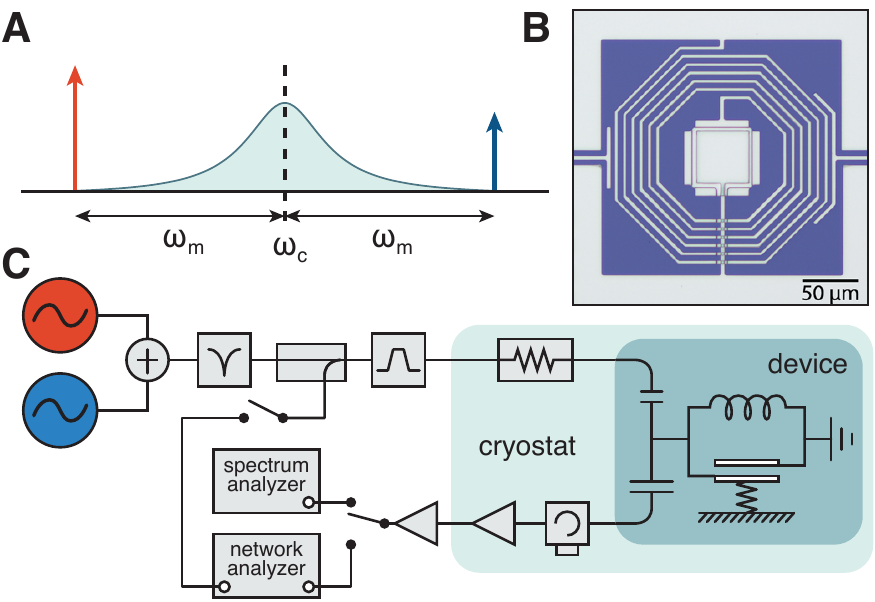}
\end{center}
\caption{Measurement set-up and device. (\textbf{A}) Schematic showing squeezing drive tones detuned by \wm~from the cavity frequency. The red-detuned tone has a greater amplitude than the blue-detuned tone. (\textbf{B}) Optical micrograph of the device. Gray regions are aluminum, while blue regions are the exposed silicon substrate. The center square is a parallel-plate capacitor with a top plate that has a mechanical degree of freedom. The capacitor is surrounded by a spiral inductor, and coupling capacitors on the left and right provide input and output coupling to the device. (\textbf{C}) Microwave circuit diagram for squeezing measurements. The red- and blue-detuned tones are filtered at room temperature and attenuated at low temperatures so that they are shot-noise limited at the device. The device is thermally-connected to the mixing plate of a dilution refrigerator. Signals are amplified and then measured with a spectrum analyzer or network analyzer.}
\label{fig:1}
\end{figure*}

\begin{figure*}[p]
\begin{center}
\includegraphics{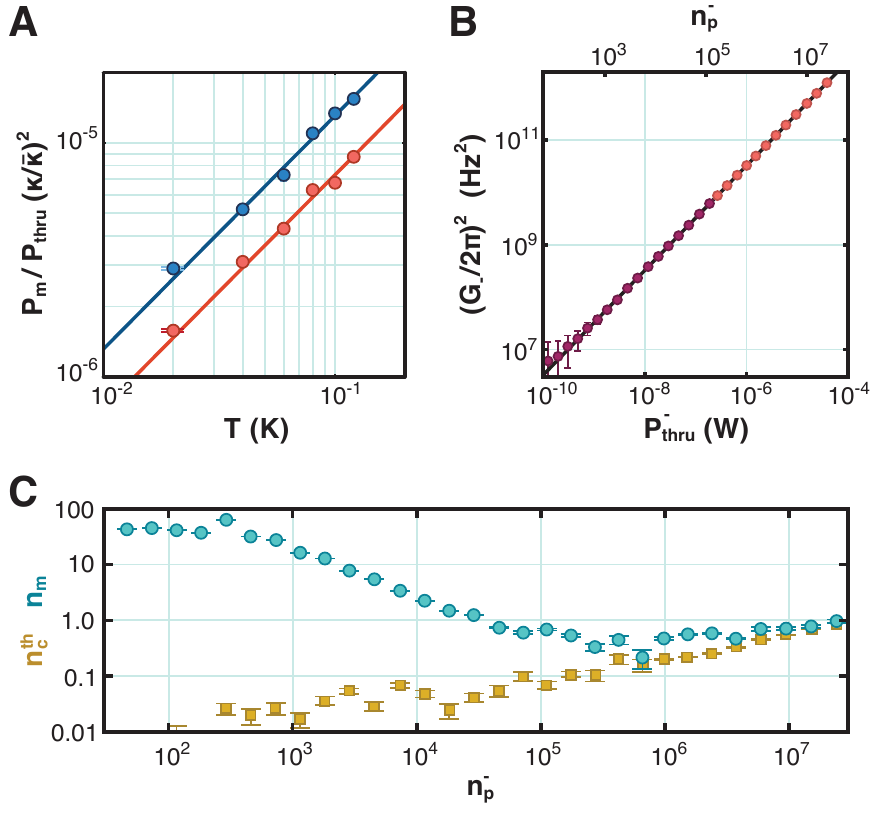}
\end{center}
\caption{Calibrations and characterization. (\textbf{A}) We measure the integrated mechanical sideband power ($P_m$) of a weak red-detuned (red) or blue-detuned (blue) pump and normalize by the measured pump power ($P_{thru}$). This quantity is proportional to the thermal occupation of the mechanics times $\go^2$. The difference in slopes is due to the difference in microwave transmission at \wred~and \wblue. (\textbf{B}) Enhanced coupling rate calibration for a red-detuned drive. Purple circles: weak-driving regime in which the mechanical susceptibility is a simple Lorentzian. Black line: fit to the weak-driving regime damping vs. power. Red circles: strong-driving regime in which the mechanical linewidth becomes comparable to the cavity linewidth, and \Gr~must be extracted using a full model. Both the linearity of our device at high pump powers and the accuracy of the model are evident in the adherence of these strong-driving points to the fit line from the weak-driving regime. (\textbf{C}) Characterization of sideband cooling in the presence of a red-detuned drive. Turquoise circles: cooled mechanical occupation. Yellow squares: cavity thermal occupation. The mechanical occupation reaches a minimum of $0.22 \pm 0.08$ quanta before heating of the cavity bath limits the occupation.}
\label{fig:2}
\end{figure*}

\begin{figure*}[p]
\begin{center}
\includegraphics{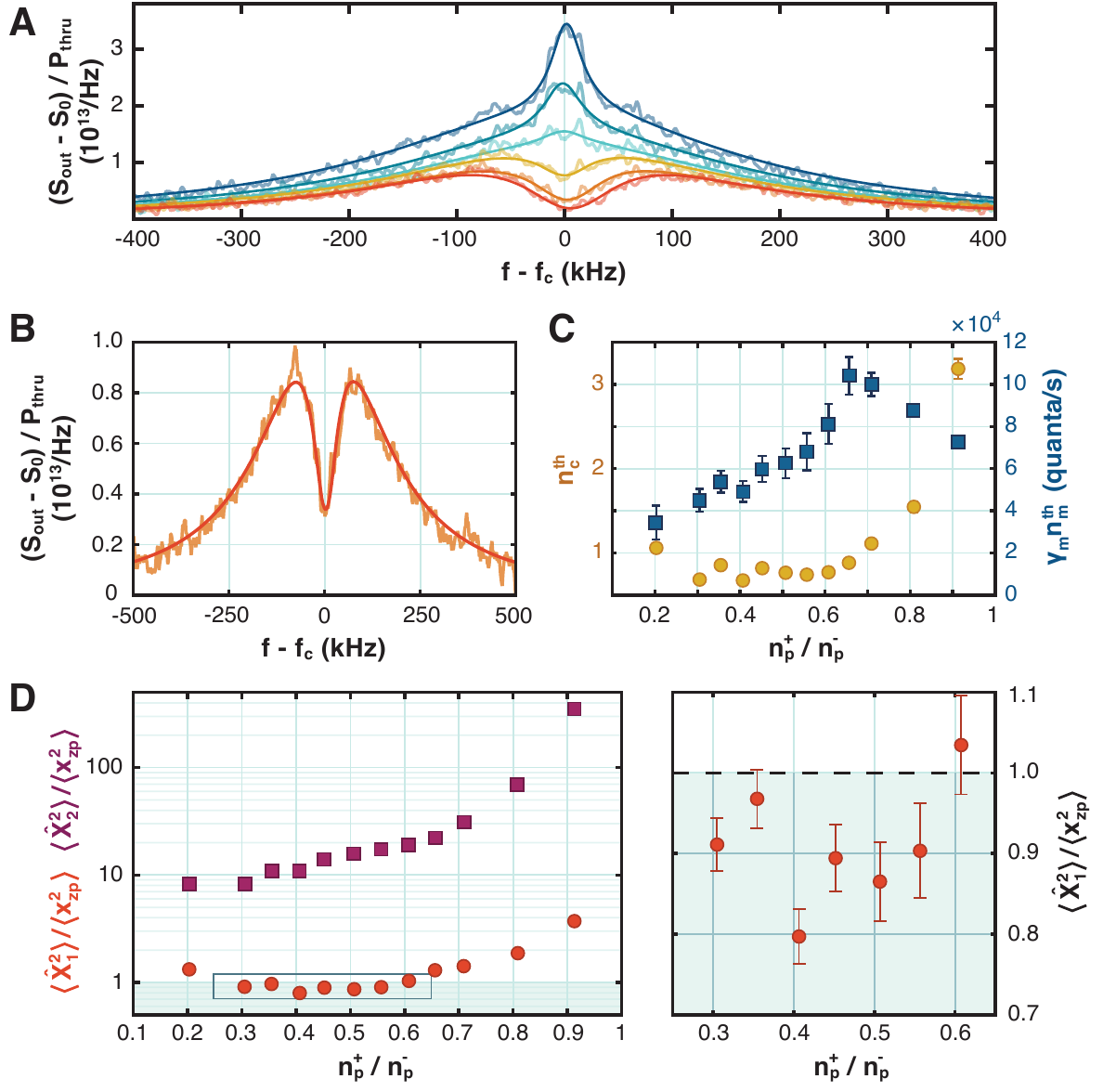}
\end{center}
\caption{Squeezing measurements. (\textbf{A}) Background-subtracted output spectra normalized by the measured red-detuned pump power at blue/red pump power ratios of 0.3, 0.4, 0.5, 0.6, 0.65, and 0.7, from red to blue. Darker lines indicate fits from Eq. \ref{eq:outspec}. All spectra are taken at the total power of $\npr+\npb = 1.76\times10^7$. As the ratio decreases, the damping increases, broadening the mechanical linewidth and causing the mechanical noise go from a peak to a dip due to noise squashing. (\textbf{B}) Close-up of spectrum and fit at $\npb/\npr = 0.4$. (\textbf{C}) Values of \ncth~(yellow circles) and \gm\nmth~(blue squares) obtained from a Bayesian analysis of the spectrum. (\textbf{D}) Left: Calculated quadrature noise for \XIsq~(red circles) and \XIIsq~(purple squares). The shaded region indicates sub-zero-point squeezing. Right: Close-up of the boxed area showing data with \XI~fluctuations below the zero-point level. The lowest point has $\XIsq/\xzpsq = 0.806 \pm 0.035$. The fit for this point is shown in (B).}
\label{fig:3}
\end{figure*}

\begin{figure}[p]
\centering
\includegraphics{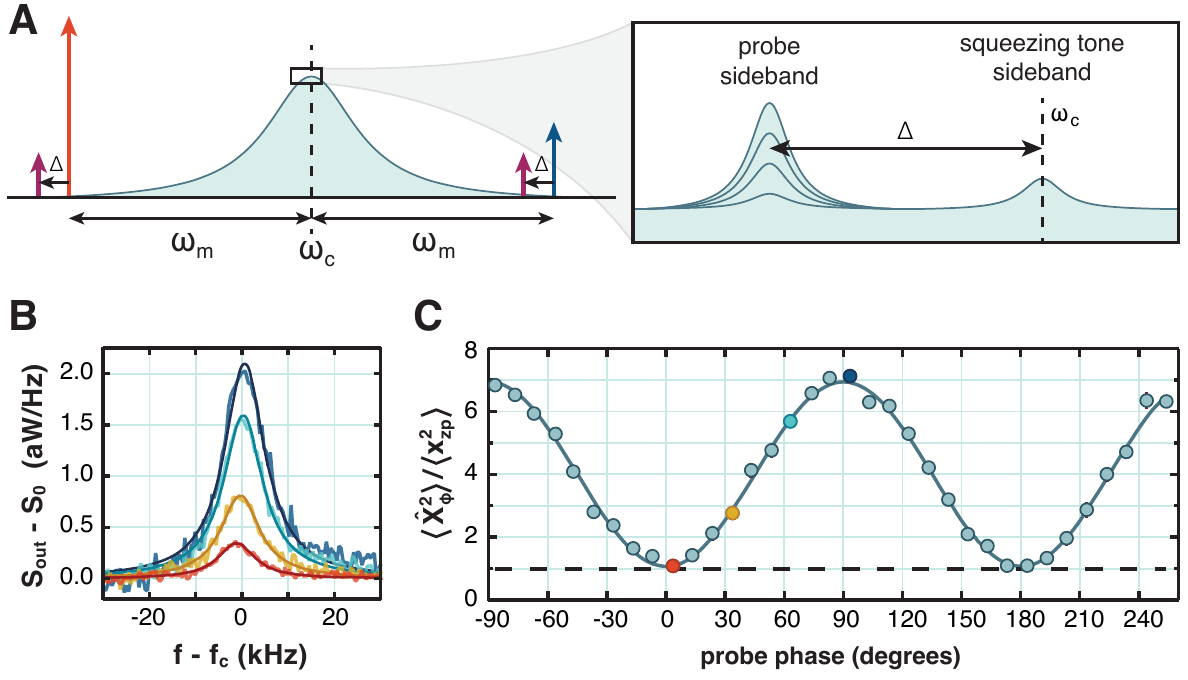}
	\caption{BAE measurement of squeezing. 
	(\textbf{A}) Schematic of pump and probe tone configuration, showing the cavity response and sideband spectra for different probe phases.
	(\textbf{B}) BAE probe sideband spectra for different probe phases with $\npr=16\times 10^6$, $\npb=3.2\times 10^6$, and each $n_p^{probe}=0.95\times10^6$.
	(\textbf{C}) Quadrature variance as a function of phase, as obtained from the area under the sideband lorentzians. The spectra shown in (B) are highlighted in their corresponding color. The zero-point variance is indicated with a dashed black line.
	\label{fig:4} }
\end{figure}

\end{document}